%% file: main.tex
\documentclass[compsoc, conference, letterpaper, 10pt, times]{IEEEtran}

\usepackage[backend=biber]{biblatex}
\usepackage{comment, graphicx, color, booktabs, algorithm, algpseudocode, xspace, listings, subcaption, multirow, enumitem, tikz, siunitx}
\usepackage[bookmarks=false, hidelinks]{hyperref}
\usepackage{eso-pic}

\usepackage[utf8]{inputenc}
\usepackage[T1]{fontenc}

\graphicspath{{figure/}}

\newcommand{\fn}[1]{\mbox{\textsc{#1}}}
\newcommand{\kw}[1]{\textbf{#1}\xspace}
\newcommand{\file}[1]{\textit{#1}\xspace}

\newcommand*{\email}[1]{\href{mailto:#1}{\nolinkurl{#1}} } 

\newcommand{\name}{Angora\xspace}

\title{\name: Efficient Fuzzing by Principled Search}

\def\preprint{preprint}
\def\mode{preprint}
\ifx\mode\preprint
\author{\IEEEauthorblockN{Peng Chen}
  \IEEEauthorblockA{ShanghaiTech University\\
    \email{chenpeng@shanghaitech.edu.cn}
  }
  \and
  \IEEEauthorblockN{Hao Chen}
  \IEEEauthorblockA{University of California, Davis \\
    \email{chen@ucdavis.edu}
  }
}
\else
\author{Anonymized for submission}
\fi

\begin{document}
\AddToShipoutPictureBG*{%
  \AtPageUpperLeft{%
    \setlength\unitlength{1in}%
    \hspace*{\dimexpr0.5\paperwidth\relax}
    \makebox(0,-0.75)[c]{To appear in the 39th IEEE Symposium on Security and Privacy, May 21--23, 2018, San Francisco, CA, USA}%
}}
\maketitle
\ifx\mode\preprint
\IEEEpeerreviewmaketitle
\fi

\input{abstract}
\input{introduction}
\input{background}
\input{design}
\input{implementation}
\input{evaluation}

\input{related}

\input{conclusion}
\input{acknowledgment}
\printbibliography
\end{document}

%% file: abstract.tex
\begin{abstract}

Fuzzing is a popular technique for finding software bugs. However, the performance of the state-of-the-art fuzzers leaves a lot to be desired. Fuzzers based on symbolic execution produce quality inputs but run slow, while fuzzers based on random mutation run fast but have difficulty producing quality inputs. We propose \name, a new mutation-based fuzzer that outperforms the state-of-the-art fuzzers by a wide margin. The main goal of \name is to increase branch coverage by solving path constraints without symbolic execution. To solve path constraints efficiently, we introduce several key techniques: scalable byte-level taint tracking, context-sensitive branch count, search based on gradient descent, and input length exploration. On the LAVA-M data set, \name found almost all the injected bugs, found more bugs than any other fuzzer that we compared with, and found eight times as many bugs as the second-best fuzzer in the program \file{who}. \name also found 103 bugs that the LAVA authors injected but could not trigger. We also tested \name on eight popular, mature open source programs. \name found 6, 52, 29, 40 and 48 new bugs in \file{file}, \file{jhead}, \file{nm}, \file{objdump} and \file{size}, respectively. We measured the coverage of \name and evaluated how its key techniques contribute to its impressive performance.

\end{abstract}

%% file: introduction.tex
\section{Introduction}

Fuzzing is a popular technique for finding software bugs. Coverage-based fuzzers face the key challenge of how to create inputs to explore program states. Some fuzzers use symbolic execution to solve path constraints~\cite{cadar2008klee, cha2012unleashing}, but symbolic execution is slow and cannot solve many types of constraints efficiently~\cite{cadar2013symbolic}. To avoid these problems, AFL uses no symbolic execution or any heavy weight program analysis~\cite{afl}. It instruments the program to observe which inputs explore new program branches, and keeps these inputs as seeds for further mutation. AFL incurs low overhead on program execution, but most of the inputs that it creates are ineffective (i.e., they fail to explore new program states) because it blindly mutates the input without taking advantage of the data flow in the program. Several fuzzers added heuristics to AFL to solve simple predicates, such as ``magic bytes''~\cite{vuzzer2017, li2017steelix}, but they cannot solve other path constraints.

We designed and implemented a fuzzer, called \name\footnote{The \name rabbit has longer, denser hair than American Fuzzy Lop. We name our fuzzer \name to signify that it has better program coverage than AFL while crediting AFL for its inspiration.}, that explores the states of a program by solving path constraints without using symbolic execution. \name tracks the unexplored branches and tries to solve the path constraints on these branches. We introduced the following techniques to solve path constraints efficiently.

\begin{table}
\caption{Bugs found on the LAVA-M data set by different fuzzers. Note that \name found more bugs than listed by LAVA authors.} 
\begin{center}
\setlength{\tabcolsep}{0.11cm} 
\begin{tabular}{@{}lSSSSSSS}
\toprule
\multirow{2}{*}{Program} & \multicolumn{1}{l}{Listed} & \multicolumn{5}{c}{Bugs found by each fuzzer}\\ \cmidrule{3-8}
& \multicolumn{1}{l}{bugs} & \multicolumn{1}{l}{\textbf{\name}} & \multicolumn{1}{l}{AFL} & \multicolumn{1}{l}{FUZZER} & \multicolumn{1}{l}{SES} & \multicolumn{1}{l}{VUzzer} & \multicolumn{1}{l}{Steelix} \\
\midrule
\file{uniq} & 28 &  29 & 9 & 7 & 0 & 27 & 7 \\
\file{base64} & 44 & 48 & 0 & 7 & 9 & 17 & 43 \\
\file{md5sum} & 57 & 57 & 0 & 2 & 0 & \multicolumn{1}{l}{Fail} & 28\\
\file{who} & 2136 & 1541 & 1 & 0 & 18 & 50 & 194 \\
\bottomrule
\end{tabular}
\end{center}
\label{tbl:lava_all}
\end{table}

\begin{itemize}

\item \emph{Context-sensitive branch coverage}. AFL uses context-insensitive branch coverage to approximate program states. Our experience shows that adding context to branch coverage allows \name to explore program states more pervasively (\autoref{sec:context_sensitive}).

\item \emph{Scalable byte-level taint tracking}. Most path constraints depend on only a few bytes in the input. By tracking which input bytes flow into each path constraint, \name mutates only these bytes instead of the entire input, therefore reducing the space of exploration substantially (\autoref{sec:tainttrack}).

\item \emph{Search based on gradient descent}. When mutating the input to satisfy a path constraint, \name avoids symbolic execution, which is expensive and cannot solve many types of constraints. Instead, \name uses the gradient descent algorithm popular in machine learning to solve path constraints (\autoref{sec:gd}).

\item \emph{Type and shape inference}. Many bytes in the input are used collectively as a single value in the program, e.g., a group of four bytes in the input used as a 32-bit signed integer in the program. To allow gradient descent to search efficiently, \name locates the above group and infers its type (\autoref{sec:infer_type}).

\item \emph{Input length exploration}. A programs may explore certain states only when the length of the input exceeds some threshold, but neither symbolic execution nor gradient descent can tell the fuzzer when to increase the length of the input. \name detects when the length of the input may affect a path constraint and then increases the input length adequately
 (\autoref{sec:len_exploration}).
\end{itemize}

\name outperformed state-of-the-art fuzzers substantially. \autoref{tbl:lava_all} compares the bugs found by \name with other fuzzers on the LAVA-M data set~\cite{lava2016}. \name found more bugs in each program in the data set. Particularly, in \file{who} \name found 1541 bugs, which is eight times as many bugs as found by the second-best fuzzer, Steelix. Moreover, \name found 103 bugs that the LAVA authors injected but could not trigger. We also tested \name on eight popular, mature open source programs. \name found 6, 52, 29, 40 and 48 new bugs in \file{file}, \file{jhead}, \file{nm}, \file{objdump} and \file{size}, respectively (\autoref{tbl:real_app}). We measured the coverage of \name and evaluated how its key techniques contribute to its impressive performance.

%% file: background.tex
\section{Background: American Fuzzy Lop (AFL)}
\label{sec:afl}
Fuzzing is an automated testing technique to find bugs. American Fuzzy Lop (AFL)~\cite{afl} is a state-of-the-art mutation-based graybox fuzzer. AFL employs light-weight compile-time instrumentation and genetic algorithms to automatically discover test cases that likely trigger new internal states in the targeted program. As a coverage-based fuzzer, AFL generates inputs to traverse different paths in the program to trigger bugs.

\subsection{Branch coverage}
\label{sec:branch_coverage}

AFL measures a path by a set of branches. During each run, AFL counts how many times each branch executes. It represents a branch as a tuple ($l_\mathrm{prev}, l_\mathrm{cur}$), where $l_\mathrm{prev}$ and $l_\mathrm{cur}$ are the IDs of the basic blocks before and after the conditional statement, respectively. AFL gets the branch coverage information by using lightweight instrumentation. The instrumentation is injected at each branch point at compile time. 
For each run, AFL allocates a \emph{path trace table} to count how many times each branch of every conditional statement executes. The index to the table is the hash of a branch, $h(l_\mathrm{prev}, l_\mathrm{cur})$, where $h$ is a hash function.

AFL also keeps a global \emph{branch coverage table} across different runs. Each entry contains an 8-bit vector that records how many times the branch executes in different runs. Each bit in this vector $b$ represents a range: $b_0, \ldots, b_7$ represent the ranges $[1]$, $[2]$, $[3]$, $[4, 7]$, $[8, 15]$, $[16, 31]$, $[32, 127]$, $[128, \infty)$, respectively. For example, if $b_3$ is set, then it indicates that there exists a run where this branch executed between 4 and 7 times, inclusively. 

AFL compares the path trace table and branch coverage table to determine, heuristically, whether a new input triggers a new internal state of the program. An input triggers a new internal state if either of the following happens:

\begin{itemize}
\item The program executes a new branch, i.e., the path trace table has an entry for this branch but the branch coverage table has no entry for this branch.
\item There exists a branch where the number of times, $n$, this branch executed in the current run is different from any previous runs. AFL determines this approximately by examining whether the bit representing the range of $n$ was set in the corresponding bit vector in the branch coverage table.
\end{itemize}

\subsection{Mutation strategies}
\label{sec:afl_mutation}
AFL applies the following mutations on the input randomly~\cite{afl_mutate}.
\begin{itemize}
\item Bit or byte flips.
\item Attempts to set ``interesting'' bytes, words, or dwords.
\item Addition or subtraction of small integers to bytes, words, or dwords.
\item Completely random single-byte sets.
\item Block deletion, block duplication via overwrite or insertion, or block memset.
\item Splice two distinct input files at a random location.
\end{itemize}

%% file: design.tex
\section{Design}
\label{sec:design}

\subsection{Overview}
\begin{algorithm}
    \begin{algorithmic}[1]
    \Function{Fuzz}{$program, seeds$}
    \State Instrument $program$ in two versions: $program_\mathrm{nt}$ (no taint tracking) and $program_\mathrm{t}$ (with taint tracking). 
    \State $branches \gets$ empty hash table \Comment{Key: an unexplored branch $b$. Value: the input that explored $b$'s sibling branch.}
    \ForAll{$input \in seeds$}
        \State $path \gets$ Run $program_\mathrm{t}(input)$
        \ForAll{unexplored branch $b$ on $path$}
        \State$branches[b] \gets input$
        \EndFor
    \EndFor
    \While {$branches \not= \emptyset$}
        \State Select $b$ from $branches$
        \While {$b$ is still unexplored}
            \State Mutate $branches[b]$ to get a new input $input'$ (\autoref{alg:fuzz_cond})

            \State Run $program_\mathrm{nt}(input')$
            \If {$input'$ explored new branches}
                \State $path' \gets$ Run $program_\mathrm{t}(input')$
                \ForAll{unexplored branch $b'$ on $path'$}
                \State $branches[b'] \gets input'$
                \EndFor
            \EndIf
            \If {$b$ was explored}
            \State $branches \gets branches - \{b\}$
            \EndIf
        \EndWhile
    \EndWhile
    \EndFunction
\end{algorithmic}
\caption{\name's fuzzing loop. Each \textbf{while} loop has a budget (maximum allowed number of iterations)}
\label{alg:overview}
\end{algorithm}

AFL and other similar fuzzers use branch coverage as the metric. However, they fail to consider the call context when calculating branch coverage. Our experience shows that without context, branch coverage would fail to explore program states adequately. Therefore, we propose context-sensitive branch coverage as the metric of coverage (\autoref{sec:context_sensitive}).

\autoref{alg:overview} shows \name's two stages: instrumentation and the fuzzing loop. During each iteration of the fuzzing loop, \name selects an unexplored branch and searches for an input that explores this branch. We introduce the following key techniques to find the input efficiently.

\begin{itemize}
\item For most conditional statements, its predicate is influenced by only a few bytes in the input, so it would be unproductive to mutate the entire input. Therefore, when exploring a branch, \name determines which input bytes flow into the corresponding predicate and focuses on mutating these bytes only (\autoref{sec:tainttrack}).

\item After determining which input bytes to mutate, \name needs to decide how to mutate them. Using random or heuristics-based mutations is unlikely to find satisfactory values efficiently. Instead, we view the path constraint on a branch as a constraint on a blackbox function over the input, and we adapt the gradient descent algorithm for solving the constraint (\autoref{sec:gd}).

\item During gradient descent, we evaluate the blackbox function over its arguments, where some arguments consist of multiple bytes. For example, when four consecutive bytes in the input that are always used together as an integer flow into a conditional statement, we ought to consider these four bytes as a single argument to the function instead of as four independent arguments. To achieve this goal, we need to infer which bytes in the input are used collectively as a single value and what the type of the value is (\autoref{sec:infer_type}).

\item It would be inadequate to only mutate bytes in the input. Some bugs are triggered only after the input is longer than a threshold, but this creates a dilemma on deciding the length of the input. If the input is too short, it may not trigger certain bugs. But if the input is too long, the program may run too slow. Most fuzzers change the length of inputs using ad hoc approaches. By contrast, \name instruments the program with code that detects when a longer input may explore new branches and that determines the minimum required length (\autoref{sec:len_exploration}).
\end{itemize}

\begin{figure*}[t]
  \centering \includegraphics[width=0.8\linewidth]{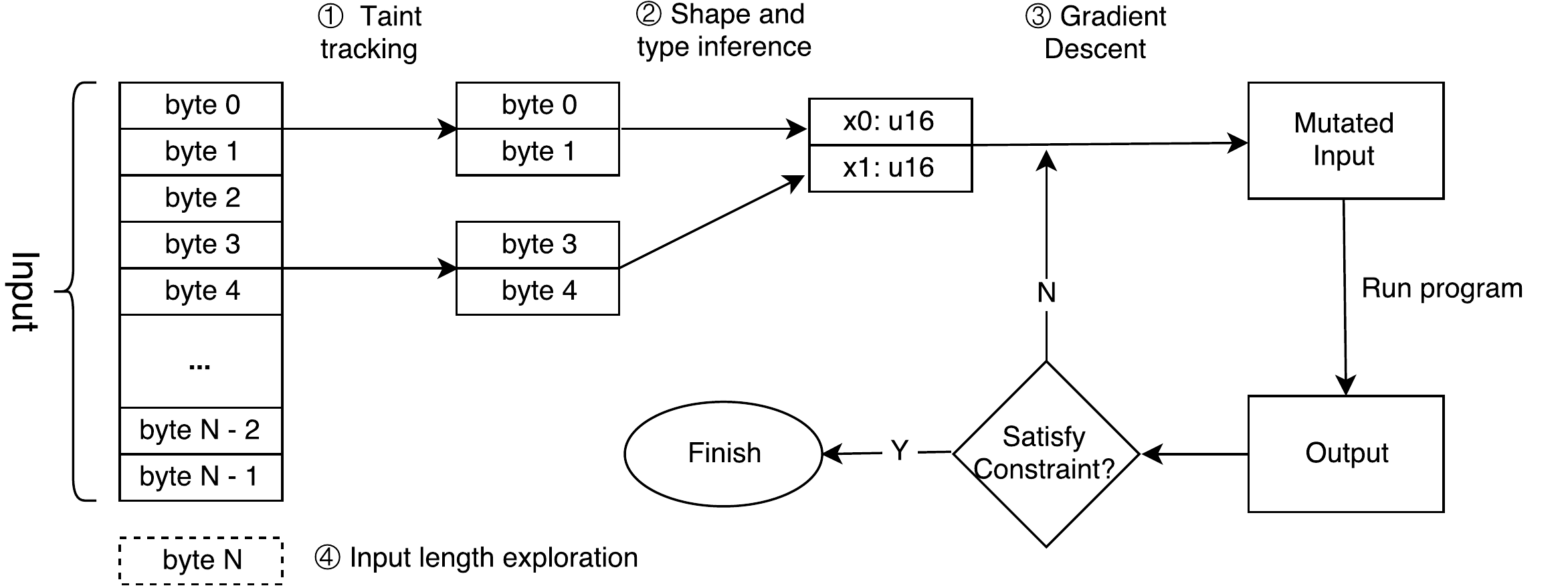}
  \caption{Steps in fuzzing a conditional statment.}
  \label{fig:overview2}
\end{figure*}

\autoref{fig:overview2} shows a diagram of the steps in fuzzing a conditional statement. The program in~\autoref{fig:main} demonstrates these steps in action.

\begin{itemize}
\item \emph{Byte-level taint tracking}: When fuzzing the conditional statement on Line~\ref{lst:main:cmp}, using byte-level taint tracking, \name determines that bytes 1024--1031 flow into this expression, so it mutates these bytes only.

\item \emph{Search algorithm based on gradient descent}: \name needs to find inputs that run both branches of the conditional statement on Line~\ref{lst:main:cmp}, respectively. \name treats the expression in the conditional statement as a function $f(\mathbf{x})$ over the input $\mathbf{x}$, and uses gradient descent to find two inputs $\mathbf{x}$ and $\mathbf{x}'$ such that $f(\mathbf{x}) > 0$ and $f(\mathbf{x}') \leq 0$.

\item \emph{Shape and type inference}: $f(\mathbf{x})$ is a function over the vector $\mathbf{x}$. During gradient descent, \name computes the partial derivative of $f$ over each component of $\mathbf{x}$ separately, so it must determine each component and its type. On Line~\ref{lst:main:cmp}, \name determines that $\mathbf{x}$ consists of two components each consisting of four bytes in the input and having the type 32-bit signed integer.

\item \emph{Input length exploration}: \texttt{main} will not call \texttt{foo} unless the input has at least 1032 bytes. Instead of blindly trying longer inputs, we instrument common functions that read from input and determine if longer input would explore new states. For example, if the initial input is shorter than 1024 bytes, then the conditional statement on Line~\ref{lst:main:fread1} will execute the true branch. Since the return value of \texttt{fread} is compared with 1024, \name knows that only inputs at least 1024 bytes long will explore the false branch. Similarly, the instrumentation on Lines~\ref{lst:main:fread2} and \ref{lst:main:fread3} instructs \name to extend the input to at least 1032 bytes to execute the function \texttt{foo}.
\end{itemize}

\begin{figure}[t]
\begin{lstlisting}
void foo(int i, int j) {
  if (i * i -  j * 2 > 0) { (*@\label{lst:main:cmp}@*)
    // some code    
  } else {
    // some code
  }
}

int main() {
  char buf[1024];
  int i = 0, j = 0;
  if(fread(buf, sizeof(char), 1024, fp) (*@\label{lst:main:fread1}@*)
    < 1024) {
    return(1);
  }
  if(fread(&i, sizeof(int), 1, fp) < 1){ (*@\label{lst:main:fread2}@*)
    return(1);
  }
  if(fread(&j, sizeof(int), 1, fp) < 1){ (*@\label{lst:main:fread3}@*)
    return(1);
  }
  foo(i, j);
}
\end{lstlisting}
\caption{Example program showing core techniques}
\label{fig:main}
\end{figure}

\subsection{Context-sensitive branch count}

\label{sec:context_sensitive}

\autoref{sec:afl} describes AFL's branch coverage table. Its design has several advantages. First, it is space efficient. The number of branches is linear in the size of the program. Second, using ranges to count branch execution provides good heuristics on whether a different execution count indicates new internal state of the program. When the execution count is small (e.g., less than four), any change in the count is significant. However, when the execution count is large (e.g., greater than 32), a change has to be large enough to be considered significant.

But this design has a limitation. Because AFL's branches are context-insensitive, they fail to distinguish the executions of the same branch in different contexts, which may overlook new internal states of the program. \autoref{fig:afl_context} illustrates this problem. Consider the coverage of the branch on Line~\ref{lst:branch}. During the first run, the program takes the input ``\texttt{10}''. When it calls \texttt{f()} on Line~\ref{lst:caller1}, it executes the true branch on Line~\ref{lst:true}. Later, when it calls \texttt{f()} on Line~\ref{lst:caller2}, it executes the false branch on Line~\ref{lst:false}. Since AFL's definition of branch is context-insensitive, it thinks that both branches have executed. Later, when the program takes a new input ``\texttt{01}'', AFL thinks that this input triggers no new internal state, since both the branches on Line~\ref{lst:true} and \ref{lst:false} executed in the previous run. But in fact this new input triggers a new internal state, as it will cause crash on Line~\ref{lst:crash} when \texttt{input[2]==1}.

\begin{figure}[t]
\begin{lstlisting}
void f(bool x) {
  static bool trigger = false;
  if (x) { (*@\label{lst:branch}@*)
    if (trigger) { (*@\label{lst:true}@*)
      if (input[2]) {
        // crash; (*@\label{lst:crash}@*)
      }
    }
  } else {
    if (!trigger) { (*@\label{lst:false}@*)
      trigger = true;
    }
  }
}

bool[] input;

int main() {
  f(input[0]); (*@\label{lst:caller1}@*)
  ...
  f(input[1]); (*@\label{lst:caller2}@*)
}
\end{lstlisting}
\caption{Example where context-insensitive branch count fails to detect new behavior}
\label{fig:afl_context}
\end{figure}

We incorporate context into the definition of branches. We define a branch as a tuple $(l_\mathrm{prev}, l_\mathrm{cur}, context)$, where $l_\mathrm{prev}$ and $l_\mathrm{cur}$ are the IDs of the basic blocks before and after the conditional statement, respectively, $context$ is $h(stack)$ where $h$ is a hash function, and $stack$ contains the state of the call stack. For example, let the program in \autoref{fig:afl_context} first run on the input \texttt{10}. After it enters \texttt{f()} from Line~\ref{lst:caller1}, it will execute the branch $(l_{\ref{lst:branch}}, l_{\ref{lst:true}}, [l_{\ref{lst:caller1}}])$. Then, after it enters \texttt{f()} from Line~\ref{lst:caller2}, it will execute the branch $(l_{\ref{lst:branch}}, l_{\ref{lst:false}}, [l_{\ref{lst:caller2}}])$. By contrast, when the program executes on the input ``\texttt{01}'', it will execute the branches $(l_{\ref{lst:branch}}, l_{\ref{lst:false}}, [l_{\ref{lst:caller1}}])$ followed by $(l_{\ref{lst:branch}}, l_{\ref{lst:true}}, [l_{\ref{lst:caller2}}])$. By incorporating calling context into the definition of branch, \name can detect that the second run triggers a new internal state, which will lead to the crash site on Line~\ref{lst:crash} when mutating \texttt{input[2]}.

Adding context to branches increases the number of unique branches, which could be dramatic when deep recursion occurs. Our current implementation mitigates this problem by selecting a particular function $h$ for computing the hash of the call stack  where $h$ computes the \textit{xor} of the IDs of all the call sites on the stack. When \name instruments the program, it assigns a random ID to each call site. Therefore, when a function \texttt{f} recursively calls itself, no matter how many times \name pushes the ID of the same call site to the call stack, $h(stack)$ outputs at most two unique values, which at most doubles the number of unique branches in the function \texttt{f}. Our evaluation on real world programs shows that after incorporating context, the number of unique branches increases by as many as 7.21 times (\autoref{tbl:context_branch_increase}) in exchange for the benefit of improved code coverage (\autoref{fig:context_compare}).

\subsection{Byte-level taint tracking}
\label{sec:tainttrack}

The objective of \name is to create inputs to execute unexplored branches. When it tries to execute an unexplored branch, it must know which byte offsets in the input affect the predicate of the branch. Therefore, \name requires byte-level taint tracking. However, taint tracking is expensive, especially when tracking each byte individually, so AFL avoids it. Our key insight is that \emph{taint tracking is unnecessary on most runs of the program}. Once we run taint tracking on an input (Step 1 in \autoref{fig:overview2}), we can record which byte offsets flow into each conditional statement. Then, when we mutate these bytes, we can run the program \emph{without} taint tracking.  This amortizes the cost of taint tracking on one input over its many mutations, which allows \name to have a similar throughput of input execution as AFL (\autoref{sec:speed}).

\name associates each variable $x$ in the program with a \emph{taint label} $t_x$, which represents the byte offsets in the input that may flow into $x$.
The data structure of taint labels has a big impact on its memory footprint. A naive implementation would be to represent each taint label as a bit vector, where each bit $i$ represents the $i$th byte in the input. However, since the size of this bit vector grows linearly in the size of the input, this data structure would be prohibitive for large input, but large input is necessary for finding bugs in certain programs.

To reduce the size of the taint labels, we could store the bit vectors in a table, and use the indices to the table as taint labels. As long as the logarithm of the number of entries in the table is much less than the length of the longest bit vectors, which is often the case, we can greatly reduce the size of the taint labels. 

However, this data structure raises a new challenge. The taint labels must support the following operations:

\begin{itemize}
\item \fn{Insert}($b$): inserts a bit vector $b$ and returns its label.
\item \fn{Find}($t$): returns the bit vector of the taint label $t$.
\item \fn{Union}($t_x, t_y$): returns the taint label representing the union of the bit vectors of the taint labels $t_x$ and $t_y$.
\end{itemize}

\fn{Find} is cheap, but \fn{union} is expensive. \fn{union} takes the following steps. First, it finds the bit vectors of the two labels and computes their union $u$. This step is cheap. Next, it searches the table to determine if $u$ already exists. If not, it adds $u$. But how to search efficiently? A linear search would be expensive. Alternatively, we could build a hash set of the bit vectors, but if there are a lot of them and each bit vector is long, it would take much time to compute the hash code and much space to store the hash set. Since \fn{union} is a common operation when we track tainted data in arithmetic expressions, it must be efficient. Note that we cannot use the \fn{union-find} data structure because the vectors are not disjoint, i.e., two different bit vectors may have 1 at the same position.

We propose a new data structure for storing the bit vectors that allows efficient \fn{insert}, \fn{find} and \fn{union}. For each bit vector, the data structure assigns it a unique label using an unsigned integer. When the program inserts a new bit vector, the data structure assigns it the next available unsigned integer.

The data structure contains two components. 

\begin{itemize}
\item A binary tree maps bit vectors to their labels. Each bit vector $b$ is represented by a unique tree node $v_b$ at level $|b|$, where $|b|$ is the length of $b$. $v_b$ stores the label of $b$. To reach $v_b$ from the root, examine $b_0, b_1, \ldots $ sequentially. If $b_i$ is 0, go to the left child; otherwise, go to the right child. Each node contains a back pointer to its parent to allow us to retrieve the bit vector starting from $v_b$.

\item A look up table maps labels to their bit vectors. A label is an index to this table, and the corresponding entry points to the tree node representing the bit vector of this label.
\end{itemize}

In this data structure, all leaves in the tree represent bit vectors, and no internal node represents bit vectors. However, many nodes in the tree may be unnecessary. For example, if a vector $x00*$ is in the tree but no vector $x0[01]*1[01]*$ is in the tree, where $x$ is any sequence of bits, then it would be unnecessary to store any node after the node representing $x$, because $x$ has only one decedent that is a leaf, and this leaf represents $x00*$. Here we use the common notation for regular expressions where $x*$ means that $x$ is repeated zero or more times, and $[xy]$ means either $x$ or $y$. This observation allows us to trim a vector when inserting it into a tree as follows:

\begin{enumerate}
\item Remove all the trailing 0s of the vector.
\item Follow the bits in the vector, from the first to the last bit, to traverse the tree.
\begin{itemize}
\item If a bit is 0, follow the left child
\item Otherwise, follow the right child.
\end{itemize}
If a child is missing, create it.
\item Store the label of the vector in the last node we visited.
\end{enumerate}

\autoref{alg:insert} describes this insert operations in detail. \autoref{alg:find} and \autoref{alg:union} describe the \fn{find} and \fn{union} operations, respectively. Note that when we create a node, initially it contains no label. Later, if this node is the last node visited when we insert a bit vector, we store the label of the bit vector in this node. With this optimization, this tree has the following properties:

\begin{itemize}
\item Each leaf node contains a label.
\item An internal node may contain a label. We may store a label in an internal node that has no label yet, but we never replace the label in any internal node.
\end{itemize}

\begin{algorithm}
\begin{algorithmic}[1]
\Function{insert}{$root, vector, nodes$}
\Comment{$root$: root of the tree. $vector$: the vector to be inserted. $nodes$: an array indexed by labels containing pointers to tree nodes. $return$: the label representing $vector$.}
\State Trims all the trailing 0s in $vector$
\If {$vector$ is empty}
    \If {$root$ contains no label}
        \State $root.label \gets nodes.length()$
        \Comment{Assigns the next available integer as the label for this vector.}
        \State $nodes.push(root)$
    \EndIf
    \State \kw{return} $root.label$
\EndIf

\If {$vector[0] == 0$}
    \State $node \gets root.left$
\Else
    \State $node \gets root.right$
\EndIf
\If {$node$ does not exist}
    \State Creates $node$
    \If {$|vector| == 1$}
        \State $node.label = nodes.length()$
        \Comment{Assigns the next available integer as the label for this vector.}
        \State $nodes.push(node)$
    \EndIf
\EndIf

\If {$|vector| == 1$}
    \State \kw{return} $node.label$
\Else
    \State \kw{return} \fn{insert}($node, vector[1..], nodes$)
    \Comment{$vector[1..]$ is $vector$ after first element removed.}
\EndIf
\EndFunction
\end{algorithmic}
\caption{Insert a bit vector into the tree}
\label{alg:insert}
\end{algorithm}

\begin{algorithm}
\begin{algorithmic}[1]
\Function{find}{$label, nodes$} 
\Comment{
$label$: an integer representing a tree node, which represents a bit vector.
$nodes$: an array indexed by labels and containing pointers to tree nodes. $return$: the bit vector represented by $label$.}
\State $vector \gets$ empty vector
\State $node \gets nodes[label]$
\State $parent \gets node.parent$
\While {$parent$ exists}
    \If {$node$ is the left child of $parent$}
        \State $vector.insert\_at\_beginning(0)$
    \Else
        \State $vector.insert\_at\_beginning(1)$
    \EndIf
    \State $node \gets parent$
    \State $parent \gets node.parent$
\EndWhile 
\State \kw{return} vector
\EndFunction
\end{algorithmic}
\caption{Find a bit vector by its label}
\label{alg:find}
\end{algorithm}

\begin{algorithm}
\begin{algorithmic}[1]
\Function{union}{$label_1, label_2, nodes, root$}
\Comment{$label_1, label_2$: labels of two bit vectors. $nodes$: a table containing pointers to tree nodes. $root$: root of the tree. Return: the label representing the union of the bit vectors.}
\State $v_1 \gets \fn{find}(label_1, nodes)$
\State $v_2 \gets \fn{find}(label_2, nodes)$
\State $v \gets v_1 \cup v_2$
\State \kw{return} \fn{insert}($root, v, nodes$)
\EndFunction
\end{algorithmic}
\caption{Union two bit vectors}
\label{alg:union}
\end{algorithm}

This data structure greatly reduces the memory footprint for storing the bit vectors. Let the length of each bit vector be $n$, and let there be $l$ bit vectors. If we naively store all the bit vectors in a look up table, it would take $O(nl)$ space. However, in our data structure, the number of nodes in the tree is $O(l)$. Each node may store at most one index to the look up table. Since the look up table has $l$ entries and each entry is a pointer and so has a fixed size, the size of the look up table is $O(l)$, and each index to the look up table has $O(\log l)$ bits. Therefore, the total space requirement is $O(l\cdot\log l)$.

\subsection{Search algorithm based on gradient descent}
\label{sec:gd}

Byte-level taint tracking discovers which byte offsets in the input flow into a conditional statement. But how to mutate the input to run the unexplored branch of the statement? Most fuzzers mutate the input randomly or using crude heuristics, but those strategies are unlikely to find an appropriate input value quickly. By contrast, we view this as a search problem and take advantage of search algorithms in machine learning. We used gradient descent in our implementation, but other search algorithms might also work.

In this approach, we view the predicate for executing a branch as a constraint on a \emph{blackbox} function $f(\mathbf{x})$, where $\mathbf{x}$ is a vector of the values in the input that flow into the predicate, and $f()$ captures the computation on the path from the start of the program to this predicate. There are three types of constraints on $f(\mathbf{x})$:

\begin{enumerate}
\item $f(\mathbf{x}) < 0$.
\item $f(\mathbf{x}) <= 0$.
\item $f(\mathbf{x}) == 0$.
\end{enumerate}

\autoref{tbl:constraint} shows that we can transform all forms of comparison into the above three types of constraints. If the predicate of a conditional statement contains logical operators \texttt{\&\&} or \texttt{||}, \name splits the statement into multiple conditional statements. For example, it splits \verb+if (a && b) { s } else { t }+ into \verb+if (a) { if (b) {s} else {t} } else {t} +.

\begin{table}
\caption{Transforming comparisons into constraints. $a$ and $b$ represent arbitrary expressions.}
\begin{center}
\begin{tabular}{lll}
\toprule
Comparison & f & Constraint \\
\midrule
$a < b$ & $f=a-b$ & $f < 0$\\
$a <= b$ & $f=a-b$ & $f <= 0$\\
$a > b$ & $f=b-a$ & $f < 0$\\
$a >= b$ & $f=b-a$ & $f <= 0$\\
$a == b$ & $f=abs(a-b)$ & $f == 0$\\
$a != b$ & $f=-abs(a-b)$ & $f < 0$\\
\bottomrule 
\end{tabular}
\end{center}
\label{tbl:constraint}
\end{table}

\autoref{alg:fuzz_cond} shows the search algorithm. Starting from an initial $\mathbf{x_0}$, find $\mathbf{x}$ such that $f(\mathbf{x})$ satisfies the constraint. Note that to satisfy each type of constraint, we need to minimize $f(\mathbf{x})$, and we use gradient descent for this purpose.

Gradient descent finds a minimum of a function $f(\mathbf{x})$. The method is iterative. Each iteration starts from an $\mathbf{x}$, computes $\nabla_\mathbf{x} f(\mathbf{x})$ (the gradient of $f(\mathbf{x})$ at $\mathbf{x}$), and updates $\mathbf{x}$ as $\mathbf{x} - \epsilon \nabla_\mathbf{x} f(\mathbf{x})$ where $\epsilon$ is the learning rate.

When training neural networks, researchers use gradient descent to find a set of weights that minimize the training error. However, gradient descent has the problem that it sometimes may be stuck in a local minimum that is not a global minimum. Fortunately, this is often not a problem in fuzzing, because we only need to find an input $\mathbf{x}$ that is good enough instead of a globally optimal $\mathbf{x}$. For example, if the constraint is $f(\mathbf{x})<0$, then we just need to find an $\mathbf{x}$ where $f(\mathbf{x})<0$ instead of where $f(\mathbf{x})$ is a global minimum.

However, we face unique challenges when applying gradient descent to fuzzing. Gradient descent requires computing the gradient $\nabla_\mathbf{x} f(\mathbf{x})$. In neural networks, we can write $\nabla_\mathbf{x} f(\mathbf{x})$ in an analytic form. However, in fuzzing, we have no analytic form of $f(\mathbf{x})$. Second, in neural networks, $f(\mathbf{x})$ is a continuous function because $\mathbf{x}$ contains the weights of the network, but in fuzzing $f(\mathbf{x})$ is usually a discrete function. This is because most variables in a typical program are discrete, so most elements in $\mathbf{x}$ are discrete.

We solve these problems using numerical approximation. The gradient of $f(\mathbf{x})$ is the unique vector field whose dot product with any unit vector $\mathbf{v}$ at each point $\mathbf{x}$ is the directional derivative of $f$ along $\mathbf{v}$. We approximate each directional derivative by $\frac{\partial f(\mathbf{x})}{\partial \mathbf{x}_i}=\frac{f(\mathbf{x}+\delta \mathbf{v}_i)-f(\mathbf{x})}{\delta}$ where $\delta$ is a small positive value (e.g., 1) and $\mathbf{v}_i$ is the unit vector in the $i$th dimension. To compute each directional derivative, we need to run the program twice, once with the original input $\mathbf{x}$ and once with the perturbed input $\mathbf{x}+\delta \mathbf{v}_i$. It is possible that in the second run, the program fails to reach the program point where $f(\mathbf{x}+\delta \mathbf{v}_i)$ is calculated because the program took a different branch at an earlier conditional statement. When this happens, we set $\delta$ to a small negative value (e.g., -1) and try to compute $f(\mathbf{x}+\delta \mathbf{v}_i)$ again. If this succeeds, we compute the directional derivative based on it. Otherwise, we set the derivative to zero, instructing gradient descent not to move $\mathbf{x}$ in this direction. The time for computing the gradient is proportional to the length of the vector $\mathbf{x}$ since \name computes each directional derivative separately. \autoref{sec:infer_type} will describe how to reduce the length of $\mathbf{x}$ by merging continuous bytes that are used as a single value in the program.

In theory gradient descent can solve any constraint. In practice, how fast gradient descent can solve a constraint depends on the complexity of the mathematical function.
\begin{itemize}
\item If $f(\mathbf{x})$ is monotonic or convex, then gradient descent can find a solution quickly even if $f(\mathbf{x})$ has a complex analytic form. For example, consider the constraint $f(\mathbf{x}) < 0$ where $f(\mathbf{x})$ approximates $\log(\mathbf{x})$ using some polynomial series. This constraint would be very difficult for symbolic execution to solve because of the complex analytic form. However, it is easy for gradient descent to solve because $f(\mathbf{x})$ is monotonic. 
\item If the local minimum that gradient descent finds satisfies the constraint, finding the solution is also quick.
\item If the local minimum does not satisfy the constraint, \name has to randomly walk to another value $\mathbf{x'}$ and start to perform gradient descent from there hoping to find another local minimum that satisfies the constraint. 
\end{itemize}
Note that \name does not produce an analytic form of $f(\mathbf{x})$ but rather runs the program to compute $f(\mathbf{x})$. 

\begin{algorithm}
\begin{algorithmic}[1]
    
\Function{FuzzConditionalStmt}{$stmt, input$}
\Comment{$stmt$: The conditional statement to fuzz. $input$: The input to the program}  
\Repeat
\State $\mathbf{grad} \gets $\fn{CalculateGradient}($stmt, input$)
\If {$\mathbf{grad} == \mathbf{0}$}
    \State $input \gets $ \fn{ReSample}($input$)
    \State{Continue to the next iteration}
\EndIf

\State $input, value \gets$ \fn{Descend}($stmt, input, grad$)

\Until{\fn{SatisfyConstraint}($stmt, value$) or timeout} 

\EndFunction

\end{algorithmic}
\caption{Using gradient descent to solve path constraints}
\label{alg:fuzz_cond}
\end{algorithm}

\subsection{Shape and type inference}
\label{sec:infer_type}

Naively, we could let each element in $\mathbf{x}$ be a byte in the input that flows into the predicate. However, this would cause problems in gradient descent because of type mismatch. For example, let the program treat four consecutive bytes $b_3b_2b_1b_0$ in the input as an integer, and let $\mathbf{x}_i$ represent this integer value. When computing $f(\mathbf{x}+\delta \mathbf{v}_i)$, we should add $\delta$ to this integer. But if we naively assign each byte $b_3$, $b_2$, $b_1$, $b_0$ to a different element in $\mathbf{x}$, then we would compute $f(\mathbf{x}+\delta \mathbf{v}_i)$ on each of these bytes, but this is inappropriate. The program combines these bytes as a single value and uses only the combined value in expressions, so when we add a small $\delta$ to any byte other than the least significant byte, we would change this combined value significantly, which would cause the calculated partial derivative to be a poor approximation of the true value.

To avoid this problem, we must determine (1) which bytes in the input are always used together as a single value in the program, and (2) what is the type of the value. We call the first problem \emph{shape inference}, the second problem \emph{type inference}, and solve them during dynamic taint analysis. For shape inference, initially all the bytes in the input are independent. During taint analysis, when an instruction reads a sequence of input bytes into a variable where the size of the sequence matches the size of a primitive type (e.g., 1, 2, 4, 8 bytes), \name tags these bytes as belonging to the same value. When conflicts arise, \name uses the smallest size. For type inference, \name relies on the semantics of the instruction that operates on the value. For example, if an instruction operates on a signed integer, then \name infers the corresponding operand to be a signed integer. When the same value is used both as signed and unsigned types, \name treats it as the unsigned type. Note that when \name fails to infer the precise size and type of a value, this does not prevent gradient descent from finding a solution --- the search just takes longer.

\begin{algorithm}
\begin{algorithmic}[1]
\Procedure{InferTypes}{}

\State $type\_table \gets Array(0, INPUT\_SIZE)$
\Comment{ An array whose size is the same as the input, and all elements are 0 initially.}

\ForAll{$inst \in $ memory\_read\_instructions}
\State $address, size \gets$ \fn{GetMemReadInfo($inst$)}
\If{\fn{IsVaildType($size$)}}

\If {\fn{IsConsecutiveBytes($address$, $size$)}}
\State $offset \gets$ \fn{GetInputOffset($address$)}
\If {$type\_table[offset] == 0 $ or \\ ~~~~~~~~~~~~~~~~~~~ $ type\_table[offset] > size$}
\State $type\_table[offset] \gets size$
\EndIf
\EndIf
\EndIf

\EndFor

\EndProcedure
\end{algorithmic}
\caption{Infer which bytes in the input are used collectively as single values in the program}
\label{alg:infertype}
\end{algorithm}

\subsection{Input length exploration}
\label{sec:len_exploration}

\name, like most other fuzzers, starts fuzzing with inputs as small as possible. However, some branches are executed only when the input is longer than a threshold. This creates a dilemma for the fuzzer. If the fuzzer uses too short inputs, it cannot explore those branches. But if it uses too long inputs, the program may run slow or even out of memory. Most tools try inputs of different lengths using ad hoc approaches. By contrast, \name increases the input length only when doing so might explore new branches. 

During taint tracking, \name associates the destination memory in the \texttt{read}-like function calls with the corresponding byte offsets in the input. It also marks return value from the \texttt{read} calls with a special label. If the return value is used in a conditional statement and the constraint is not satisfied, \name increases the input length so that the \texttt{read} call can get all the bytes that it requests. For example, in \autoref{fig:main}, if the conditional statement is false on Line~\ref{lst:main:fread1}, \name extends the input length so that \texttt{fread} can read all the 1024 bytes that it requests. Our criteria are not exhaustive because programs could consume the input and check its length in ways that we have not anticipated, but it would be easy to add those criteria to \name once we discover them.

%% file: implementation.tex
\section{Implementation}
\subsection{Instrumentation}
For each program to be fuzzed, \name produces corresponding executables by instrumenting the program with LLVM Pass~\cite{llvm}. The instrumentation
\begin{itemize}
\item collects basic information of conditional statements, and links a conditional statement to its corresponding input byte offsets with taint analysis. On each input, \name runs this step only once (not while mutating this input).
\item records execution traces to identify new inputs.
\item supports context at runtime (\autoref{sec:context_sensitive}).
\item gathers expression values in predicates (\autoref{sec:gd}).
\end{itemize}

To support scalable byte-level taint tracking described in \autoref{sec:tainttrack}, we implemented taint tracking for \name by extending DataFlowSanitizer (DFSan)~\cite{dfsan}. We implemented caching facility for operations \fn{find} and \fn{union}, which speeds up taint tracking significantly .

\name depends on LLVM 4.0.0 (including DFSan). Its LLVM pass has 820 lines of C++ code excluding DFSan, and the runtime has 1950 lines of C++ code, including the data structure for storing taint labels and the hooks for tainting the input and tracking conditional statements.

In addition to the \texttt{if} statement, which has two branches, LLVM IR also supports the \texttt{switch} statement, which may introduce multiple branches. In our implementation, \name translates each \texttt{switch} statement to a sequence of \texttt{if} statements for convenience.

\name recognizes \file{libc} functions for comparing strings and arrays when they appear in conditional statements. For example, \name transforms ``\texttt{strcmp(x, y)}'' into ``\texttt{x strcmp y}'', where \texttt{strcmp} is a special comparison operator understood by \name.

\subsection{Fuzzer}
We implemented \name in 4488 lines of Rust code. We optimized \name with techniques such as \texttt{fork server}~\cite{afl_technique} and CPU binding.

%% file: evaluation.tex
\section{Evaluation}
\label{sec:evaluation}

We evaluated \name in three steps. First, we compared the performance of \name with other state-of-the-art fuzzers. Then, we measured the test coverage of \name and its ability to find unknown bugs in real world programs. Finally, we evaluated its key novel features. 

We ran all our experiments on a server with an Intel Xeon E5-2630 v3 and 256 GB memory running 64-bit Ubuntu 16.04 LTS. Even though \name can fuzz a program on multiple cores simultaneously, we configured it to fuzz the program on only one core during evaluation to compare its performance with other fuzzers. We ran each experiment five times and report the average performance.

\subsection{Compare \name with other fuzzers}
\label{sec:comparison}

The ultimate metric for comparing fuzzers is their ability to find bugs. A good test set should contain real programs with realistic bugs. LAVA is a technique for producing ground-truth corpora by injecting a large number of realistic bugs into program source code~\cite{lava2016}. The authors created a corpus LAVA-M by injecting multiple bugs into each program. LAVA-M consists of four GNU \file{coreutils} programs: \file{uniq}, \file{base64}, \file{md5sum}, and \file{who}. Each injected bug has an unique ID, which is printed when the bug is triggered.

We compared \name with the following state-of-the-art fuzzers:

\begin{itemize}
\item FUZZER (a coverage-based fuzzer) and SES (symbolic execution and SAT solving). The LAVA authors ran both of them for five hours~\cite{lava2016}.

\item VUzzer: a fuzzer using the ``magic bytes'' strategy~\cite{vuzzer2017}. Its authors reported the number of bugs found in the programs in LAVA-M, but not the running time.

\item Steelix: a fuzzer outperforming VUzzer on LAVA-M~\cite{li2017steelix}. The authors reported the number of bugs found in the programs in LAVA-M by running the fuzzer for five hours.

\item AFL 2.51b: the latest version of AFL as of this writing. We ran AFL for five hours, where we provided AFL with one CPU core for fuzzing each program. \footnote{An author of LAVA mentioned some compilation issues of running AFL on LAVA in his blog post~\cite{lava_post}, and we fixed these issues in our evaluation.}
  
\item \name: We used the same set up (one CPU core per program) as AFL.
  
\end{itemize}

\autoref{tbl:lava_all} compares the bugs found by all the fuzzers. AFL performed the worst, finding a total of 10 bugs in all the programs. VUzzer's authors could not run it on \file{md5sum} because the LAVA authors incorrectly modified \file{md5sum} to cause it to crash on all the inputs. We confirmed this problem with the LAVA authors and fixed it. Steelix is the second best fuzzer, finding almost all the bugs in \file{base64}, but only 7 out of 28 injected bugs in \file{uniq}, 28 out of 57 injected bugs in \file{md5sum}, and 194 out of 2136 injected bugs in \file{who}. \name outperformed Steelix by a large margin, finding all the bugs in \file{uniq}, \file{base64}, and \file{md5sum}, and 1443 out of 2136 injected bugs in \file{who}.

LAVA assigns each injected bug a unique ID, which is printed when the bug is triggered. The file \file{validated\_bugs} lists all the injected bugs that the LAVA authors were able to trigger when creating LAVA. \name found not only all the listed bugs in \file{uniq}, \file{base64}, \file{md5sum} and most listed bugs in \file{who}, but also 103 unlisted bugs (bugs that the LAVA authors injected but were unable to trigger). \autoref{tbl:lava_unlisted} shows the IDs of these unlisted bugs.  \autoref{tbl:lava_name} shows the breakdown of the listed and unlisted bugs found by \name.

\begin{table}[t] 
\caption{IDs of bugs injected but unlisted by LAVA, because the LAVA authors were unable to trigger them when preparing the data set. \name found these bugs.} 
\begin{center}
\begin{tabular}{lp{0.75\linewidth}}
\toprule
Program & IDs of bugs unlisted by LAVA-M but found by \name \\
\midrule
\file{uniq} & 227\\
\file{base64} & 274, 521, 526, 527 \\
\file{md5sum} & - \\
\file{who} & 2, 4, 6, 8, 12, 16, 24, 55, 57, 59, 61, 63, 73, 77, 81, 85, 89, 125, 165, 169, 173, 177, 181, 185, 189, 193, 197, 210, 214, 218, 222, 226, 294, 298, 303, 307, 312, 316, 321, 325, 327, 334, 336, 338, 350, 359, 468, 472, 477, 481, 488, 514, 526, 535, 974, 975, 995, 1007, 1026, 1034, 1071, 1072, 1415, 1429, 1436, 1456, 1718, 1735, 1736, 1737, 1738, 1747, 1748, 1755, 1756, 1891, 1892, 1893, 1894, 1903, 1904, 1911, 1912, 1921, 1925, 1935, 1936, 1943, 1944, 1949, 1953, 2231, 3264, 3545, 3551, 3939, 4287, 4295\\
\bottomrule
\end{tabular}
\end{center}
\label{tbl:lava_unlisted}
\end{table}

\begin{table}[t] 
\caption{Bugs found by \name and the corresponding running time on the LAVA-M data set. Listed bugs are in LAVA's \file{validated\_bugs} file. Unlisted bugs were not triggered when LAVA's authors prepared the data set.}
\begin{center}
\begin{tabular}{lSSSS}
\toprule
\multirow{2}{*}{Program} & \multicolumn{1}{l}{Listed} & \multicolumn{2}{c}{Found bugs} & {\multirow{2}{*}{Time (min)}} \\
\cmidrule{3-4}
& \multicolumn{1}{l}{bugs} & \multicolumn{1}{l}{Listed} & \multicolumn{1}{l}{Unlisted} & \\
\midrule
\file{uniq} & 28 & 28 & 1  & 10\\
\file{base64} & 44 & 44 & 4 & 10\\
\file{md5sum} & 57 & 57 & 0 & 10\\
\file{who} & 2136 & 1443 & 98 & 45\\
\bottomrule
\end{tabular}
\end{center}
\label{tbl:lava_name}
\end{table}

\begin{figure}[t]
  \centering \includegraphics[width=1\linewidth]{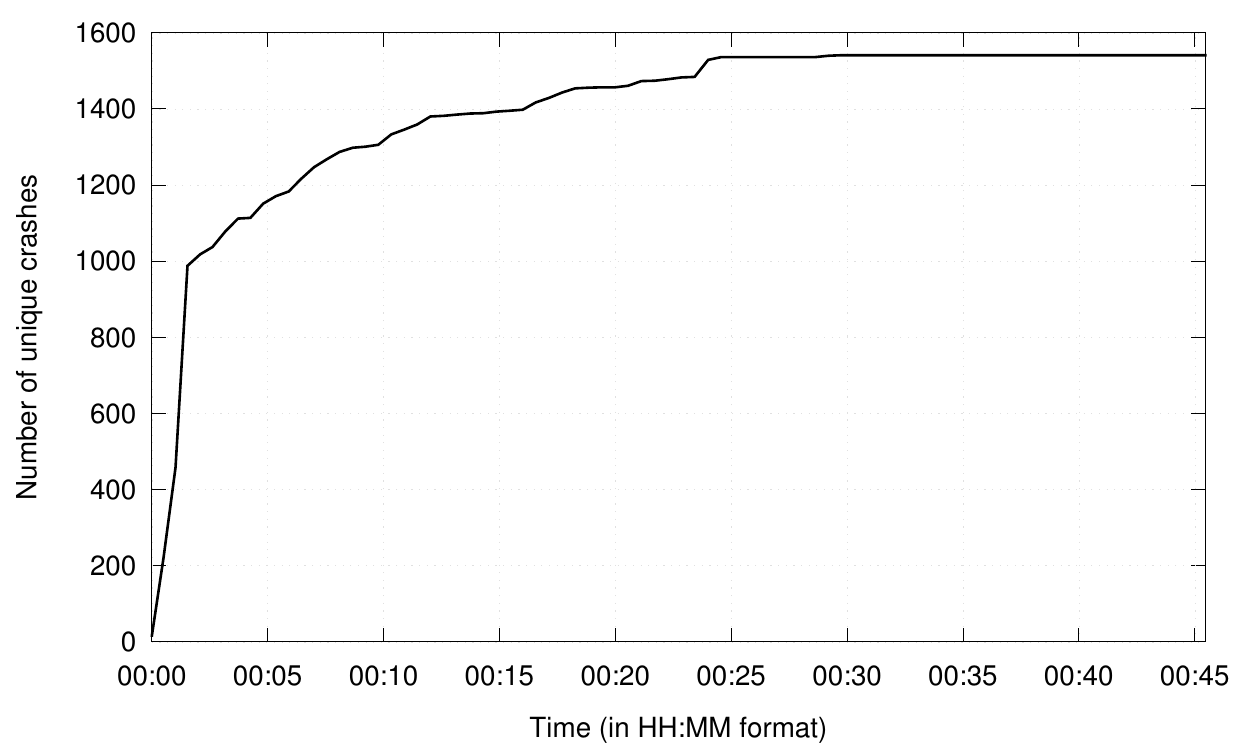}
  \caption{Cumulative number of bugs in \file{who} found by \name over time}
\label{fig:lava_who}
\end{figure}

\autoref{fig:lava_who} shows the cumulative number of bugs in \file{who} found by \name over time. We did not show the results by the other fuzzers because they found few bugs in \file{who}. \autoref{fig:lava_who} shows that initially \name discovered bugs quickly, finding 1000 bugs in less than five minutes. Then the discovery rate slowed, but it still found more than 1500 bugs in merely 45 minutes, out of the total 2136 listed bugs.

We explain why \name found a magnitude more bugs than the next best fuzzer as follows. First, LAVA uses ``magic bytes'' to guard branches that contain bugs, but some magic bytes are not copied from the input directly but rather are computed from the input. Since VUzzer and Steelix's ``magic bytes'' strategy can only copy magic bytes to the input directly, that strategy cannot create inputs that explore those branches. By contrast, \name tracks the input byte offsets that flow into a predicate, and then mutates these offsets by gradient descent instead of assuming ``magic bytes'' or any other special relation between the input and the predicate, so \name can find inputs that explore those branches. Second, VUzzer tries the ``magic bytes'' strategy blindly, and Steelix focuses on the ``magic bytes'' strategy once one of the magic bytes  matches a byte in the input after a random mutation. By contrast, \name schedules all its computing power to solve path constraints on unexplored branches, so it can cover more branches and therefore find most of the injected bugs in LAVA-M quickly.

\subsection{Evaluate \name on unmodified real world programs}
\label{sec:real_app_dataset}

\name has impressive performance on LAVA, finding not only most of the listed bugs but also many unlisted bugs. However, its skeptic might contend that these bugs were artificially injected. To address this concern, we evaluated \name on eight popular open source programs using their latest versions. Since these mature, popular programs had been extensively tested, we expected them to have few residue crashing bugs. Therefore, besides measuring the number of new bugs found, we also measured \name's coverage on these programs. We used \file{gcov}, which records all the lines and branches executed in a program on an input~\cite{gcov}. We fed each input generated by \name to the program compiled with \file{gcov} to obtain the cumulative code coverage, and \file{afl-cov}\footnote{\name is compatible with \file{afl-cov}}allowed us to do this automatically. We also ran AFL on these programs for comparison. \autoref{tbl:real_app} shows the results after running \name and AFL with one CPU core for five hours, respectively. We deduplicated the crashes by AFL's \texttt{afl-cmin -C} command.

\begin{figure*}[t]
\centering
\begin{subfigure}[b]{0.48\linewidth}
  \includegraphics[width=\textwidth]{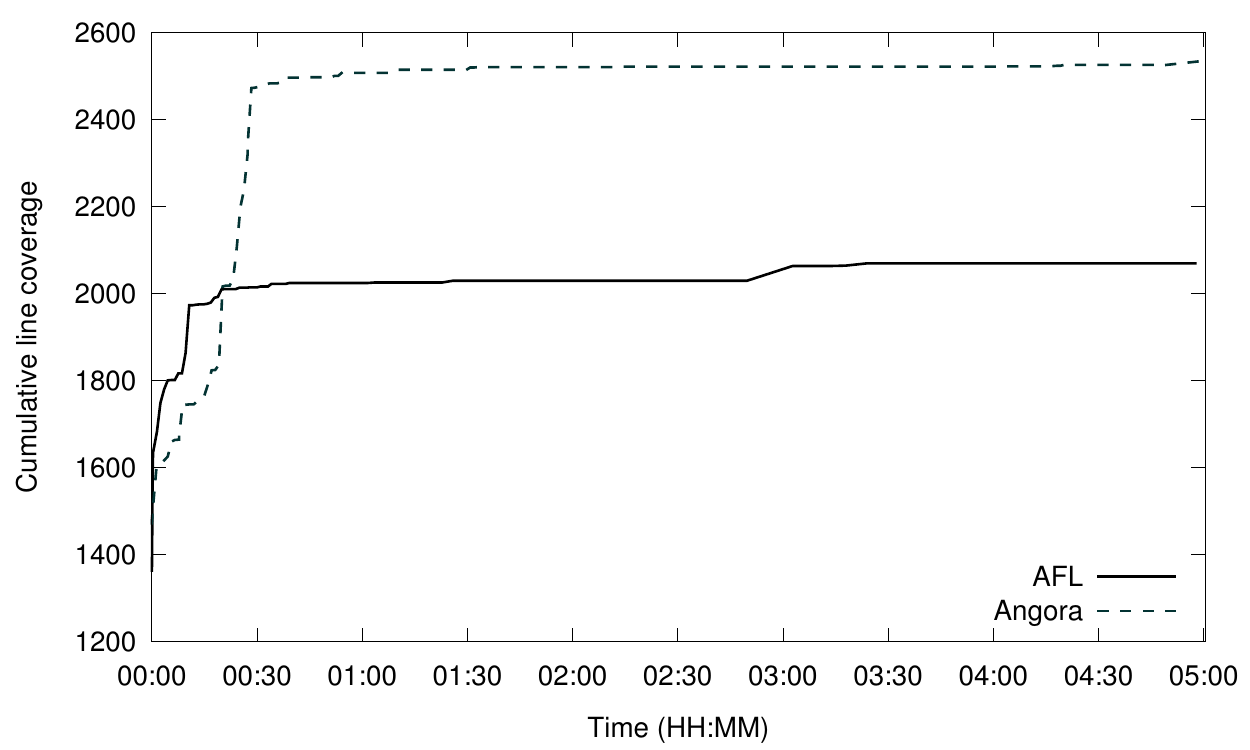}
  \caption{Line coverage}
\end{subfigure}
\begin{subfigure}[b]{0.48\linewidth}
  \includegraphics[width=\textwidth]{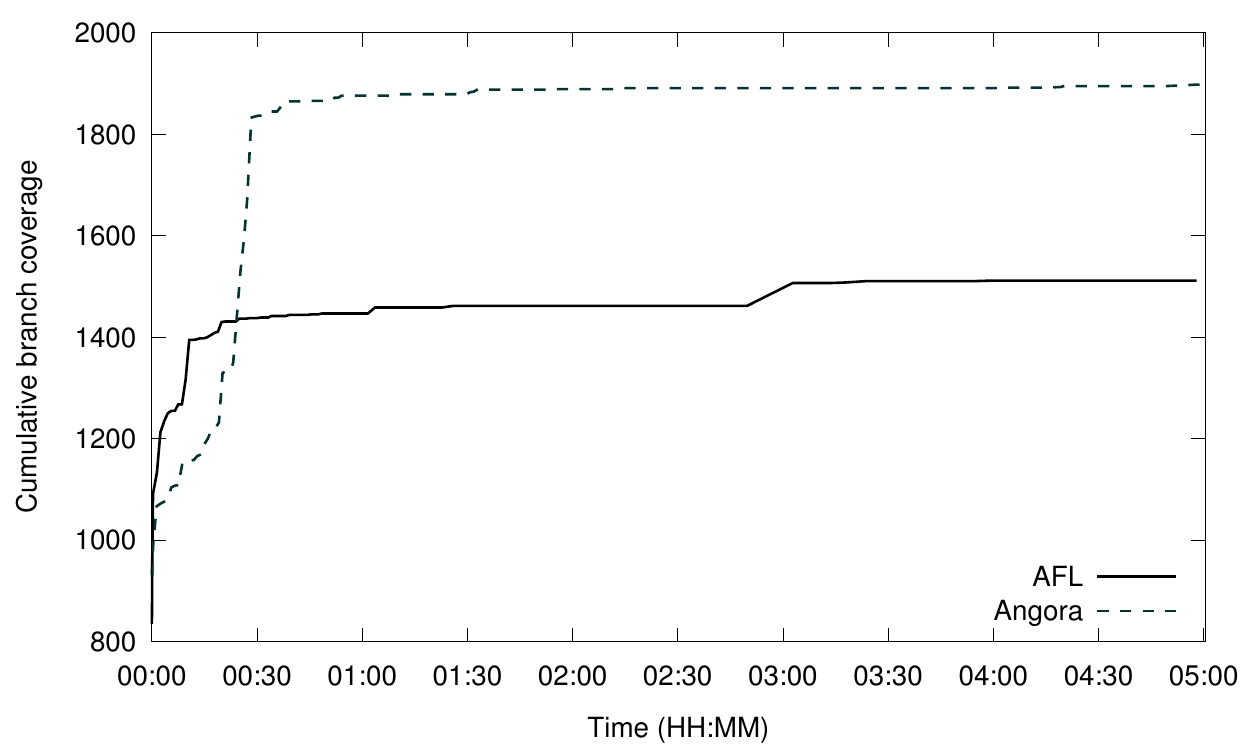}
  \caption{Branch coverage}
\end{subfigure}
\caption{Line and branch coverage on \file{file} by \name and AFL in five hours}
\label{fig:file_cov}
\end{figure*}

\autoref{tbl:real_app} shows that \name outperformed AFL on line coverage, branch coverage, and found crashes on each program. In \file{file}, \file{jhead}, \file{nm}, \file{objdump}, and \file{size}, AFL found 0, 19, 12, 4, 6 unique crashes while \name found 6, 52, 29, 40 and 48 unique crashes, respectively. The contrast is the most prominent on \file{jhead}, where \name improved the line coverage by 127.4\%, and branch coverage by 144.0\%. \autoref{fig:file_cov} compares the cumulative line and branch coverage by \name and AFL over time. It shows that \name covers more lines and branches than AFL at all time. The reason for \name's superior coverage is that it can explore both branches of complicated conditional statements. For example, \autoref{fig:complicated_cond_in_afl} shows such a statement in \file{file}, where \name successfully explored both branches but AFL could not explore the true branch.

\begin{table*}[t] 
\caption{Comparison of \name and AFL on real world programs}
\begin{center}
\begin{tabular}{llSSSrSSrSS}
  \toprule
  \multirow{2}{*}{Program} & \multirow{2}{*}{Argument} & \multicolumn{1}{l}{Size} & \multicolumn{3}{c}{Line coverage} & \multicolumn{3}{c}{Branch coverage} & \multicolumn{2}{l}{Unique crashes} \\ \cmidrule{4-11}
   & & \multicolumn{1}{l}{(kB)} & \multicolumn{1}{l}{AFL} & \multicolumn{1}{l}{\name} & \multicolumn{1}{l}{Increase} & \multicolumn{1}{l}{AFL} & \multicolumn{1}{l}{\name} & \multicolumn{1}{l}{Increase} & \multicolumn{1}{l}{AFL} & \multicolumn{1}{l}{\name} \\
\midrule
\file{file-5.32}                & & 617  & 2070 & 2534 &  \SI{21.2}{\percent}  & 1462 & 1899 & \SI{29.9}{\percent}  & 0  & 6  \\
\file{jhead-3.00}               & & 120 & 347  & 789  & \SI{127.4}{\percent} & 218  & 532  & \SI{144.0}{\percent} & 19 & 52 \\
\file{xmlwf(expat)-2.2.5}       & & 791 & 1980 & 2025 & \SI{2.3}{\percent}   & 2905 & 3158 & \SI{8.7}{\percent}   & 0  & 0  \\
\file{djpeg(ijg)-v9b}           & & 790 & 5401 & 5509 & \SI{2.0}{\percent}   & 1677 & 1782 & \SI{6.3}{\percent}   & 0  & 0  \\
\file{readpng(libpng)-1.6.34}   & & 972 & 1592 & 1799 & \SI{13.0}{\percent}  & 872  & 1007 & \SI{15.5}{\percent}  & 0  & 0  \\
\file{nm-2.29}                  &\texttt{-C} &  6252  & 6372 & 7721 & \SI{21.2}{\percent}  & 4105 & 4693 & \SI{14.3}{\percent}  & 12 & 29 \\
\file{objdump-2.29}             &\texttt{-x} & 9063 & 3448 & 6216 & \SI{80.3}{\percent}  & 2071 & 3393 & \SI{63.8}{\percent}  & 4  & 40 \\
\file{size-2.29}                & &  6207   & 2839 & 4832 & \SI{70.2}{\percent}  & 1792 & 2727 & \SI{52.2}{\percent}  & 6  & 48 \\
\bottomrule
\end{tabular}
\end{center}
\label{tbl:real_app}
\end{table*}

\begin{figure}[t]
\begin{lstlisting}[numbers=none, xleftmargin=0.5em]
// readelf.c:620
if (namesz == 10 &&
  strcmp((char*)&nbuf[noff], "DragonFly")==0
  && type == NT_DRAGONFLY_VERSION
  && descsz == 4) {
  ...
}
\end{lstlisting}
\caption{A complicated conditional statement in the file \file{readelf.c} in the program \file{file} whose true branch \name explored successfully but AFL could not}
\label{fig:complicated_cond_in_afl}
\end{figure}

In the next sections, we will evaluate how each of \name's key features contributes to its superior performance.

\subsection{Context-sensitive branch count}
\label{sec:context_evaluation}

\subsubsection{Performance}

\begin{figure}[t]
  \centering \includegraphics[width=1\linewidth]{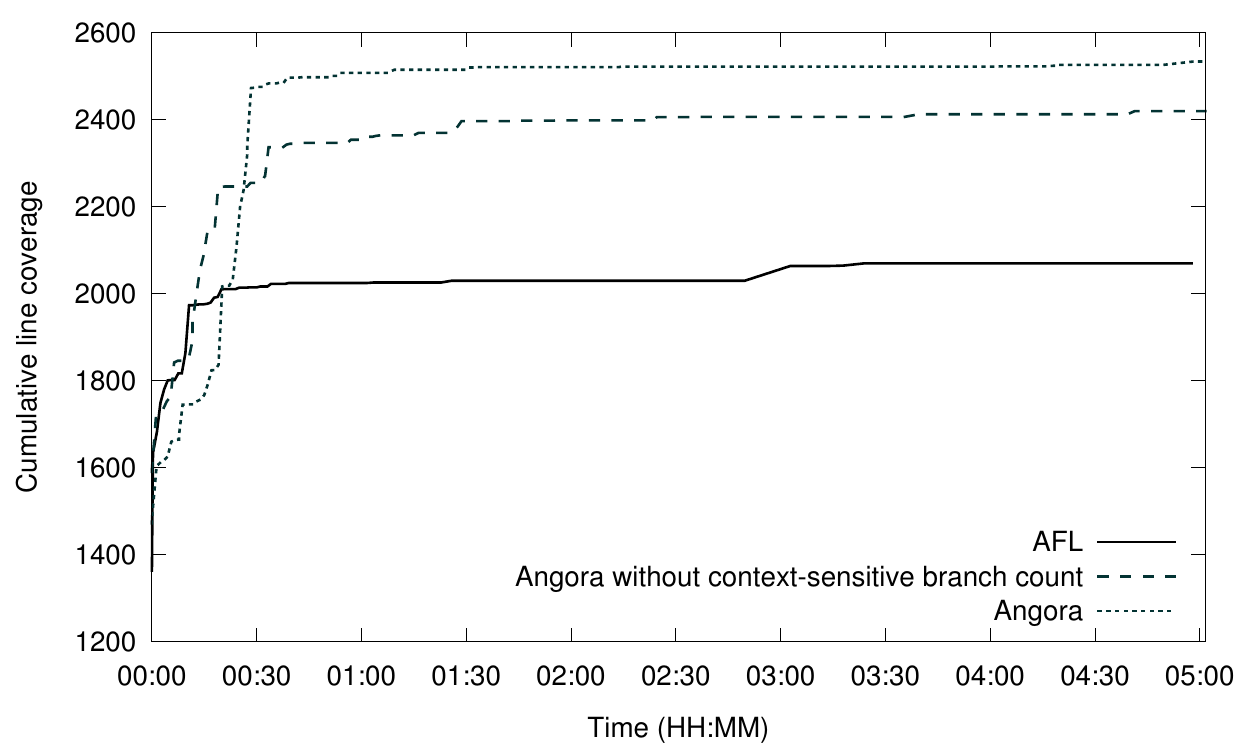}
  \caption{Comparison of non-context-sensitive vs.\ context-sensitive branch count on line coverage on the program \file{file} in five hours}
  \label{fig:context_compare}
\end{figure}

\autoref{sec:context_sensitive} introduced context-sensitive branch count. We believe that distinguishing the same branch in different function call contexts will find more bugs. To evaluate this hypothesis, we ran \name on \file{file} with context-sensitive branch count and context-insensitive branch count separately. \autoref{tbl:context_compare} shows that \name found 6 bugs with context-sensitive branch count, but no bug without it. \autoref{fig:context_compare} shows that starting from 30 minutes into fuzzing, \name consistently covered more cumulative lines with context-sensitive branch count. We discovered several real world examples where context-sensitive branch count allowed \name to explore more paths. For example, \autoref{fig:context_case_in_file} shows a code snippet in the file \file{readelf.c} in the program \file{file}. The function \texttt{getu32} is called in multiple contexts, and it returns different results based on the \texttt{swap} argument. Without context-sensitive branch count, \name would not be able to explore both branches of the conditional statement in all calling contexts.

\begin{table}[t] 
\caption{Comparison of non-context-sensitive branch count vs.\  context-sensitive branch count on the program \file{file}}
\begin{center}
\begin{tabular}{lSS}
  \toprule
Metric & \multicolumn{1}{c}{Non-context-sensitive} & \multicolumn{1}{c}{Context-sensitive}  \\
\midrule                                          
  Line coverage & 2416 & 2534 \\
  Branch coverage  & 1788 & 1899 \\
  Unique crashes & 0 & 6 \\
\bottomrule
\end{tabular}
\end{center}
\label{tbl:context_compare}
\end{table}

\begin{figure}[t]
\begin{lstlisting}[numbers=none,xleftmargin=0.5em]
// readelf.c:96
uint32_t getu32(int swap, uint32_t value) {
  ...
  if (swap) {
    ...
    return retval.ui;
  } else
    return value;
}
\end{lstlisting}
\caption{An example showing that without context sensitive branch count, \name would not be able to explore both branches of the conditional statement, because \texttt{getu32} is called from different contexts}
\label{fig:context_case_in_file}
\end{figure}

\subsubsection{Hash collision}
\label{sec:context_collision}

Similar to AFL, \name stores branch counts in a hash table. When \name incorporates calling context when counting branch coverage, it will insert more unique branches into the hash table, so we have to increase the size of the hash table to keep the collision rate low. We evaluated how many more unique branches context sensitivity brings on the real-world programs described in \autoref{sec:real_app_dataset}. The author of AFL observed that the number of unique branches (without context) usually ranges between 2k and 10k, and a hash table with $2^{16}$ buckets should be enough for common cases~\cite{afl_technique}. \autoref{tbl:context_branch_increase} shows that incorporating context sensitivity increases the number of unique branches by a factor of at most 8, which requires us to increase the size of the hash table by also a factor of 8 to have the same expected hash collision rate. By default \name allocates $2^{20}$ buckets in its hash table, which is 16 times as large as the hash table in AFL and should be adequate for most programs. Although growing the hash table may be harmful when it no longer fits in the cache, unlike AFL, which traverses the hash table both to find new paths and to prioritize inputs that cover many basic blocks, for each input \name traverses the hash table only once to find new paths. Therefore, \name is less affected by the growing size of the hash table, as demonstrated by the execution speed in \autoref{sec:speed}). 

\begin{table}[t]
\caption{Impact of incorporating context sensitivity on unique branches}
\begin{center}
\begin{tabular}{lSSS}
\toprule
\multirow{2}{*}{Program} & \multicolumn{2}{c}{Unique branches} & {\multirow{2}{*}{Ratio}}\\
\cmidrule{2-3}
& \multicolumn{1}{l}{Context insensitive} & \multicolumn{1}{l}{Context sensitive} & \\ 
  \midrule
\file{file}    & 3578 & 13554 & 3.79 \\
\file{jhead}   & 1049 & 6914  & 6.59 \\ 
\file{xmlwf}   & 5531 & 11746 & 2.20 \\ 
\file{djpeg}   & 3819 & 13787 & 3.61 \\ 
\file{readpng} & 7130 & 27577 & 3.87 \\ 
\file{nm}      & 9029 & 65131 & 7.21 \\ 
\file{objdump} & 8113 & 40539 & 5.00 \\ 
\file{size}    & 5964 & 40148 & 6.73 \\ 
\bottomrule
\end{tabular}
\end{center}
\label{tbl:context_branch_increase}
\end{table}

\subsection{Search based on gradient descent}
\autoref{sec:gd} described how to use gradient descent to solve constraints in conditional statements. We compared gradient descent with two other strategies: random mutation, and VUzzer's magic bytes plus random mutation. To exclude other variables in the measurement, we ensure that the three strategies receive the same inputs: we collected the inputs generated by AFL in \autoref{sec:real_app_dataset}, and fed them to \name as the only inputs to fuzz. We ran \name for two hours using the above three strategies respectively.

\autoref{tbl:mut_method_fair} shows that gradient descent solved more constraints than the other two strategies on all the programs. As explained in the last paragraph of \autoref{sec:comparison}, the ``magic bytes'' strategy cannot solve constraints whose values are not copied directly from the input. For example, the variable \texttt{descsz} in \autoref{fig:complicated_cond_in_afl} is used in many constraints in the program, but it is not copied from the input directly, so the ``magic bytes'' strategy did not help.

\begin{table}[t] 
\caption{Percentage of solved constraints in conditional statements using three strategies}
\begin{center}
\begin{tabular}{lSSS}
  \toprule

\multirow{2}{*}{Program} & {\multirow{2}{*}{Random}} & \multicolumn{1}{c}{Magic bytes} &  {\multirow{2}{*}{Gradient descent}}\\
& & \multicolumn{1}{c}{+ random} &  \\ 
\midrule
\file{file}    & \SI{63.5}{\percent} & \SI{76.0}{\percent} & \SI{87.1}{\percent} \\ 
\file{jhead}   & \SI{86.9}{\percent} & \SI{87.1}{\percent} & \SI{97.6}{\percent} \\ 
\file{xmlwf}   & \SI{77.4}{\percent} & \SI{81.4}{\percent} & \SI{97.0}{\percent} \\ 
\file{djpeg}   & \SI{66.1}{\percent} & \SI{73.6}{\percent} & \SI{78.3}{\percent} \\ 
\file{readpng} & \SI{19.9}{\percent} & \SI{23.7}{\percent} & \SI{24.5}{\percent} \\ 
\file{nm}      & \SI{57.5}{\percent} & \SI{66.4}{\percent} & \SI{80.2}{\percent} \\ 
\file{objdump} & \SI{47.0}{\percent} & \SI{54.9}{\percent} & \SI{56.3}{\percent} \\ 
\file{size}    & \SI{44.1}{\percent} & \SI{52.4}{\percent} & \SI{54.3}{\percent} \\ 
\bottomrule
\end{tabular}
\end{center}
\label{tbl:mut_method_fair}
\end{table}

\subsection{Input length exploration}

\begin{table}[t] 
  \caption{Comparison of \name's input length exploration vs. other tools' random strategy. The \emph{total} columns report how many times the strategies created a longer input, respectively. The \emph{useful} columns report how many of these inputs successfully explored new branches, respectively. The two rightmost columns report the average lengths of the inputs in the \emph{useful} columns, respectively.}
  \begin{center}
  \begin{tabular}{@{}lrrrrrr@{}}
    \toprule

    \multirow{3}{*}{Program} & \multicolumn{4}{c}{Longer inputs} & \multicolumn{2}{c}{Average length} \\ \cmidrule{2-7}
    & \multicolumn{2}{c}{Random} & \multicolumn{2}{c}{\name}
    & \multirow{2}{*}{Random} & \multirow{2}{*}{\name}  \\ 
\cmidrule{2-3} \cmidrule{4-5}
& Useful & Total & Useful & Total & \\
  \midrule
  \file{file}    & 185 & 79k  & 251 & 3342  & 889.9  & 399.0  \\
  \file{jhead}   & 0 & 66k    & 0 & 26      & 0.0      & 0.0  \\
  \file{xmlwf}   & 277 & 143k & 588 & 2196  & 190.3  & 128.9  \\
  \file{djpeg}   & 32 & 106k  & 474 & 3476  & 846.6  & 283.6  \\
  \file{readpng} & 46 & 35k   & 43 & 152    & 2242.7 & 363.1  \\
  \file{nm}      & 17 & 170k  & 19 & 872    & 771.7  & 248.0  \\
  \file{objdump} & 44 & 214k  & 60 & 1614   & 1271.6 & 496.0  \\
  \file{size}    & 27 & 197k  & 33 & 1482   & 1584.5 & 949.7 \\
  \bottomrule
  \end{tabular}
  \end{center}
  \label{tbl:input_length_exploration}
  \end{table}

\autoref{sec:len_exploration} describes that \name increases the length of the input on demand when it observes that a path constraint may depend on the length, while AFL and related fuzzers increase the input length randomly. We compared these two strategies based on two criteria:

\begin{itemize}
\item How many times does the strategy increase the input length? Among the inputs created by this strategy, how many are useful? An input is useful if it explores a new branch either directly or after some mutation.

\item What is the average length of those useful inputs?
\end{itemize}

We ran \name with our proposed strategy and the random strategy for five hours respectively. \autoref{tbl:input_length_exploration} shows that \name's strategy increased the input length about two orders of magnitude fewer times than the random strategy, but it found more useful inputs in all cases except two: on \texttt{readpng} it found three fewer useful inputs out of a total of 46, and on \texttt{jhead} neither strategy found any useful input because \file{jhead} only parses the header of an image and therefore is not affected by the length of the image data. \autoref{tbl:input_length_exploration} also shows that while \name's strategy generated more useful inputs, it generated shorter inputs on average on each program tested. Shorter inputs make many programs run faster. This evaluation shows that \name's strategy generates higher quality inputs than the random strategy.

\subsection{Execution speed}
\label{sec:speed}

\name's taint tracking is expensive. However, \name runs taint tracking once for each input, and then mutates the input and runs the program many times without taint tracking, so the one-time cost is amortized. Since branch count dominates the running time of the instrumented code without taint tracking, the \name-instrumented program runs at about the same speed as its AFL-instrumented version. \autoref{tbl:speed_comparison} shows that AFL executes inputs at a slightly higher rate than \name. However, because \name generates higher-quality inputs that more likely explore new branches, \name had much better coverage and found significantly more bugs as shown earlier.

\begin{table}[t] 
\caption{Inputs tested per second}
\begin{center}
\begin{tabular}{lrr}
  \toprule

Program & AFL & \name  \\
\midrule
\file{file}    & 971.17  & 791.73  \\
\file{jhead}   & 2684.45 & 2648.91 \\
\file{xmlwf}   & 2225.07 & 2206.24 \\
\file{djpeg}   & 1439.94 & 1185.52 \\
\file{readpng} & 3374.43 & 2881.72 \\
\file{nm}      & 1633.72 & 1045.35 \\
\file{objdump} & 1882.05 & 1192.04 \\
\file{size}    & 1671.95 & 1174.55 \\
\bottomrule
\end{tabular}
\end{center}
\label{tbl:speed_comparison}
\end{table}

%% file: related.tex
\section{Related work}

\subsection{Prioritize seed inputs}

An important optimization for mutation-based fuzzers is to select the seed input wisely. Rebert \emph{et.al.}~\cite{rebert2014optimizing} formulated and reasoned about the seed selection scheduling problem. They designed and evaluated six different seed selection algorithms based on PeachFuzzer~\cite{peach_fuzzer}. The algorithms used different features to minimize the seed input set, such as execution time and file size. The result showed that heuristics employed by seed selection algorithms performed better than fully random sampling. AFLFast~\cite{bohme2016coverage} observed that most fuzzing tests exercised the same few ``high frequency'' paths. They used Markov chain to identify ``low-frequency'' paths. AFLFast prioritized the inputs that contain such path. VUzzer~\cite{vuzzer2017} used control-flow features to model a path to prioritize the input whose path is hard-to-reach. Additionally, VUzzer detected error-handing basic-blocks, and prioritized the valid inputs that do not contain these basic-blocks. By contrast, \name selects the inputs whose paths contain conditional statements with unexplored branches. This is a more general strategy, which automatically directs \name to focus on the low-frequency paths after exploring the high-frequency ones.

\subsection{Taint-based fuzzing}

Taint tracking has many uses, such as analyzing malware behavior~\cite{portokalidis2006argos}, detecting and preventing information leaks~\cite{enck2014taintdroid, sun2016taintart}, and debugging software~\cite{masri2004detecting, ganai2012dtam}. It can also be used in fuzzing. Taint-based fuzzers analyze how an application processes an input to determine which part of the input should be modified. Some of these fuzzers~\cite{ganesh2009taint, bekrar2012taint, haller2013dowsing} aimed to locate the values used in security sensitive code in input files, and then fuzzed these parts of input file to trigger crashes. For example, BuzzFuzz~\cite{ganesh2009taint} used taint tracking to find which input bytes were processed by ``attack point'' that they defined. Dowser~\cite{haller2013dowsing} considered code that likely leads to buffer overflow as security sensitive code. In other words, these fuzzers aimed to exploit bugs in the reachable paths. Woo et al.\ mentioned the trade off between exploration vs.\ exploitation~\cite{woo2013scheduling}. \name can incorporate these techniques to exploit the explored paths. Taintscope~\cite{wang2010taintscope} used taint analysis to infer checksum-handling code and bypassed these checks by control flow alteration, because these checks are hard to satisfy by mutating the input.

VUzzer~\cite{vuzzer2017} is an application-aware fuzzer that used taint analysis to locate the position of ``magic bytes'' in input files, and then assigned these magic bytes to fixed positions in the input. VUzzer can find magic bytes only when they appear continuously in the input. Steelix~\cite{li2017steelix} improved VUzzer by learning from program state where the magic bytes are located in the input and how to mutate the input to match the magic bytes efficiently. By contrast, \name applies byte-level taint tracking to get the byte offsets in the input that flow into each conditional statement, and then mutates these bytes to satisfy the condition for the unexplored branch, so \name can find many more types of values efficiently than magic bytes, e.g., non-continuous magic bytes or magic bytes that are not copied directly from the input but are computed from the input. Besides, VUzzer uses a compressed bit-set data structure to represent taint labels where each bit corresponds to a unique byte offset in the input. Therefore, the size of the taint label is large for values with a complex pattern of input byte offsets because they can not be effectively compressed. By contrast, \name stores the byte offsets in a tree and uses indices into the tree as taint labels, so the size of the taint label is constant regardless of how many input byte offsets are in the label. For example, when the taint labels of several values have the same byte offsets, VUzzer repeatedly stores these byte offsets in each taint label, but \name stores these byte offsets only once in the tree, thus greatly reducing the memory consumption. 

\name's data structure for efficiently representing taint labels is similar to reduced ordered binary decision diagrams (roBDD). roBDD was used to represent dynamic slices~\cite{zhang2004efficient} and data lineage~\cite{lin2008convicting} compactly, but to the best of our knowledge, \name is the first to use this idea to represent taint labels efficiently.

\subsection{Symbolic-assisted fuzzing}

Dynamic symbolic execution provides high semantic insight into the target application. Since such techniques know how to trigger desired program state, they can be used to find vulnerabilities in programs directly. Classic approaches performed symbolic execution to maximize code coverage to find crashes~\cite{cadar2008klee, cha2012unleashing}. But the challenges of path explosion and constraint solving make symbolic execution hard to scale~\cite{cadar2013symbolic, shoshitaishvili2016sok}. Some tools tried to mitigate this obstacle by combining it with fuzzing~\cite{godefroid2005dart, godefroid2008automated, cha2015program_adaptive, stephens2016driller}. DART~\cite{godefroid2005dart} and SAGE~\cite{godefroid2008automated} used a dynamic symbolic execution engine to modify input in fuzzing. SYMFUZZ~\cite{cha2015program_adaptive} leveraged symbolic analysis on an execution trace to detect dependencies among the bit positions in an input, and then used this dependency to compute an optimal mutation ratio to guide fuzzing. Driller~\cite{stephens2016driller} used dynamic symbolic execution only when fuzzing with AFL got stuck. However, all of them inherited the scalability problem from symbolic execution. By contrast, \name does not use symbolic execution, and can find many bugs on large programs efficiently.

%% file: conclusion.tex
\section{Conclusion}

We designed and implemented \name, a powerful mutation-based fuzzer that produces high quality inputs, thanks to the following key techniques: scalable byte-level taint tracking, context-sensitive branch count, search algorithm based on gradient descent, shape and type inference, and input length exploration. \name outperformed other state-of-the-art fuzzers by a wide margin. It found significantly more bugs than other fuzzers on LAVA-M, found 103 bugs that the LAVA authors could not trigger when they prepared the data set, and a total of 175 new bugs in eight popular, mature open source programs. Our evaluation shows that \name raised the bar of fuzzing to a new level.

%% file: acknowledgment.tex
\section{Acknowledgment}

We thank Dongyu Meng for helpful discussions throughout this project and for reviewing drafts of this paper. The paper improved substantially thanks to the detailed feedback from the anonymous reviewers.